\documentclass[11pt,a4paper]{article}
\usepackage{epsfig}
\usepackage[T1]{fontenc}    
\usepackage{graphics}
\usepackage{graphicx}
\usepackage{pstricks,pst-coil,pst-fill,pst-plot}
\usepackage[fleqn]{amsmath}    
\usepackage{amssymb}    
\usepackage{amsfonts}   
\usepackage{verbatim}   
\usepackage{mathrsfs}   
\usepackage{dsfont}
\usepackage{euscript}
\usepackage{yfonts}
\usepackage{txfonts}
\usepackage{marvosym}
\usepackage{vmargin}        

\setmarginsrb{1.8cm}{2cm}{1.8cm}{2cm}{1cm}{1cm}{1cm}{1.6cm}
 \makeatletter
 \@addtoreset{equation}{section}
 \makeatother
 
\providecommand{\bysame}{\leavevmode\hbox to3em{\hrulefill}\thinspace}
\let\tend=\rightarrow

\newtheorem{theorem}{Theorem}[section]
\newtheorem{prop}{Proposition}[section]

\newtheorem{lemme}{Lemma}[section]

\def\Proof{\medskip\noindent {\it Proof --- \ }}

\def\qed{\hfill\rule{2mm}{2mm}}

\newcommand\beq{\begin{equation}}
\newcommand\enq{\end{equation}}
\newcommand\bem{\begin{multline}}
\newcommand\enm{\end{multline}}

\def\beqa{\begin{eqnarray}}
\def\eeqa{\end{eqnarray}}
\def\ba{\begin{array}}
\def\ea{\end{array}}

\def\det{\operatorname{det}}

\newcommand{\f}[2]{{\ensuremath{%
    \mathchoice%
    {\dfrac{#1}{#2}}
    {\dfrac{#1}{#2}}
    {\frac{#1}{#2}}
    {\frac{#1}{#2}}
}}}
\newcommand{\tf}[2]{\ensuremath{#1/#2}}
\newcommand{\pa}[1]{\ensuremath{\left(#1\right)}}

\newcommand{\pac}[1]{\ensuremath{\left[#1\right]}}

\def\a{\alpha}

\def\Ga{\Gamma}
\def\de{\delta}

\def\De{\Delta}

\def\veps{\varepsilon}
\def\la{\lambda}

\def\om{\omega}

\newcommand{\mc}[1]{\ensuremath{\mathcal{#1}}}

\newcommand{\msc}[1]{\ensuremath{\mathscr{#1}}}

\newcommand{\ov}[1]{\ensuremath{\overline{#1}}}
\newcommand{\wt}[1]{\ensuremath{\widetilde{#1}}}
\newcommand{\wh}[1]{\ensuremath{\widehat{#1}}}

\newcommand{\Int}[2]{\ensuremath{\int\limits_{#1}^{#2}}}
\newcommand{\Oint}[2]{\ensuremath{\oint\limits_{#1}^{#2}}}

\newcommand{\sul}[2]{\ensuremath{\sum\limits_{#1}^{#2}}}
\newcommand{\pl}[2]{\ensuremath{\prod\limits_{#1}^{#2}}}

\newcommand{\R}{\ensuremath{\mathbb{R}}}
\newcommand{\Cx}{\ensuremath{\mathbb{C}}}

\newcommand{\Dp}[1]{\ensuremath{\partial_{#1}}}

\newcommand{\limit}[2]{\ensuremath{\underset{#1 \tend #2}{\longrightarrow} }}

\newcommand{\ex}[1]{\ensuremath{\e{e}^{#1}}}

\newcommand{\ddet}[2]{\ensuremath{\det_{#1}\pac{#2}}}

\newcommand{\abs}[1]{\ensuremath{\left| #1 \right|}}

\newcommand{\norm}[1]{\ensuremath{\left\|#1\right\|}}

\newcommand{\dd}{\mathrm{d}}
\newcommand{\e}[1]{\ensuremath{\mathrm{#1}}}

\newcommand{\intff}[2]{\ensuremath{\left [ \, #1 \,; #2 \, \right ] }}
\newcommand{\intfo}[2]{\ensuremath{\left [ \, #1 \,; #2 \, \right ) }}

\newcommand{\intoo}[2]{\ensuremath{\left ( \, #1 \,; #2 \, \right ) }}

\begin{document}

\begin{flushright}

\end{flushright}
\par \vskip .1in \noindent

\vspace{14pt}

\begin{center}
\begin{LARGE}
{\bf Low-$T$ asymptotic expansion of the solution to the Yang-Yang equation.}
\end{LARGE}

\vspace{30pt}

\begin{large}

{\bf K.~K.~Kozlowski}\footnote[1]{Universit\'{e} de Bourgogne, Institut  de Math\'{e}matiques de Bourgogne, UMR 5584 du CNRS,  France, karol.kozlowski@u-bourgogne.fr},~~
\par

\end{large}

\vspace{40pt}

\centerline{\bf Abstract} \vspace{1cm}
\parbox{12cm}{\small 
We prove that the unique solution to the Yang-Yang equation arising in the context of the thermodynamics
of the so-called non-linear Schr\"{o}dinger model admits a low-temperature expansion to all orders. 
Our approach provides a rigorous justification, for a certain class of non-linear integral equations, of the 
low-temperature asymptotic expansions that were argued previously 
in various works related to the low-temperature behaviour of integrable models.}

\vspace{10 pt}


\vspace{5pt}


\end{center}

\vspace{40pt}

\section*{Introduction\label{INT}}

The non-linear Schr\"{o}dinger model refers to a quantum field theory-based model for interacting bosons that is equivalent,
in each sector with a fixed number of particles, to a one-dimensional gas of bosons interacting through a $\de$-function potential
with strength $c$. The so-called impenetrable Bose gas which corresponds to $c=+\infty$ was introduced and solved by Girardeau 
\cite{GirardeauIntroductionImpBosons}.  Then Lieb and Liniger \cite{LiebLinigerCBAForDeltaBoseGas} 
introduced the model at general $c$ and used the so-called coordinate Bethe Ansatz 
so as to construct eigenfunctions and eigenvalues associated with the underlying spectral problem. 
The completeness of their construction was however only proven much later by Dorlas \cite{DorlasOrthogonalityAndCompletenessNLSE}. 
It was recently recovered within the discreet Laplacian approach proposed recently by van Diejen \cite{DiejenDiagonalizationIntDiscretBoseGasAndProofCompleteness}.

A method for studying the thermodynamics of the model was proposed in the seminal paper of Yang and Yang 
\cite{Yang-YangNLSEThermodynamics} in 1969. These authors gave an integral representation for the free energy $f$ of the 
non-linear Schr\"{o}dinger model 
\beq
f \; = \; \Int{ \R }{} \ln \big[ 1\;+ \;  \ex{- \f{ \veps(\la)}{ T } } \big] \cdot \f{ \dd \la  }{ 2\pi } \;.  
\label{ecriture forme energie libre}
\enq
The above representation is valid for any value $0<c \leq +\infty$ of the coupling constant
just as any value of the chemical potential $h$. The dependence on the coupling constant and the chemical
potential is entirely contained in the function $\veps$ which corresponds to the unique solution to 
the non-linear integral equation
\beq
\veps(\la) = \la^2 - h - \f{T}{2\pi} \Int{ \R }{} K(\la-\mu) \ln \big[ 1+ \ex{-\f{\veps(\mu)}{T}}\big] \cdot \dd \mu  \qquad \e{with} \qquad 
K(\la) = \f{ 2c }{ \la^2  \; + \; c^2 } \;. 
\label{ecriture eqn YY}
\enq
We shall refer to the above equation as the Yang-Yang equation. 

The method proposed  by Yang and Yang underwent additional developments which allowed 
its implementation to other integrable models 
such as the XXZ spin-$\tf{1}{2}$ spin chain.
For the latter model, the thermodynamics were characterized by a solution to a hierarchy of non-linear integral equations  
\cite{GaudinTBAXXZMassiveInfiniteSetNLIE,TakahashiTBAforXXZFiniteTinfiniteNbrNLIE}. 
Later, in the nineties, there appeared an alternative 
method \cite{KomaIntroductionQTM6VertexForThermodynamicsOfXXX,KomaIntroductionQTM6VertexForThermodynamicsOfXXZ}  
for characterizing the thermodynamics of spin chains. It allowed one to express the 
free energy of a spin chain in terms of the Trotter limit of the largest eigenvalue of the so-called
quantum transfer matrix. In the case of the XYZ model, the latter eigenvalue was shown to be fully characterized in terms of a solution to 
a single non-linear integral equation \cite{KlumperNLIEfromQTMDescrThermoXYZOneUnknownFcton} quite similar to the Yang-Yang equation
\eqref{ecriture eqn YY}. These further works on the thermodynamics of integrable models 
clearly point out the Yang-Yang equation as a prototypical form of a non-linear integral equation 
arising in the study of thermodynamics of integrable models. 
Even if the resulting representations for the free energies, much in the spirit of 
\eqref{ecriture forme energie libre}, may appear implicit, they still allow one to obtain the explicit behaviour 
of thermodynamic quantities in the low-temperature regime. For this one first has to compute 
the low-$T$ asymptotic behaviour of the solution to the Yang-Yang equation (or its analogue for other models)
and then build on it so as to extract the asymptotic behaviour of the integral representation for the free energy
such as, eg \eqref{ecriture forme energie libre}. Such calculations were considered by numerous authors and for 
many different models \cite{BortzLowTExpansionAtNonZeroMagField,
EsslerFrahmGohmanKlumperKorepinOneDimensionalHubbardModel, KlumperNLIEfromQTMDescrThermoXYZOneUnknownFcton, 
KlumperScheerenSomeLowTComputationsXXX,KozMailletSlaLowTLimitNLSE}. 

The developments mentioned in the previous paragraph were not rigorous. For instance, the method set in '69
by Yang and Yang was formal and was brought to a satisfactory level of rigour only 20 years later by Dorlas, Lewis and Pul\'{e}
\cite{DorlasLewisPuleRigorousProofYangYangThermoEqnNLSE} within the framework of large deviations. 
Likewise, the works 
\cite{BortzLowTExpansionAtNonZeroMagField,
EsslerFrahmGohmanKlumperKorepinOneDimensionalHubbardModel, KlumperNLIEfromQTMDescrThermoXYZOneUnknownFcton, 
KlumperScheerenSomeLowTComputationsXXX,KozMailletSlaLowTLimitNLSE} solely proposed 
 consistent algorithms 
allowing one to compute, order by order, the terms arising in the low-$T$ asymptotic expansion of the solution
to non-linear integral equations of Yang-Yang type. 
However such algorithms always rely on the \textit{a priori} hypothesis that the low-temperature asymptotic expansion
exists in the first place and that the remainders remain stable under various operations. 
In this paper, we address the question of building a rigorous approach to proving the existence of the 
low-temperature asymptotic expansion of the solution to \eqref{ecriture eqn YY}. 
The main result of this paper is:

\begin{theorem}
\label{Theorem existence of AE for YangYang energy}
Given a positive chemical potential $h>0$ and a positive coupling constant $c>0$ for any integer $\ell \in \mathbb{N}$, the unique solution 
to the Yang-Yang equation admits the low-T asymptotic expansion 
\beq
\veps(\la) \; = \;  \sul{k=0}{\ell} \veps_{2k}(\la) \,  T^{2k} \; + \; \e{O}(T^{2\ell+2}) \;. 
\label{Theorem ecriture DA lowT veps}
\enq
Moreover, the functions $\veps_{2k}$ are even and holomorphic in the open strip $\mc{S}_{\tf{c}{2}}$ 
where 
\beq
\mc{S}_{a} \; = \; \Big\{  z \in \Cx \; : \; | \Im(z)  |  < a   \Big\} \; .
\label{definition strip width a}
\enq
Furthermore, the remainder in \eqref{Theorem ecriture DA lowT veps} 
holds uniformly in $\la \in \mc{S}_{a}$ -for any $0<a<\tf{c}{2}$- and is stable in respect to a finite order $\la$-differentiation. 
\end{theorem}

The paper is organized as follows. In Section \ref{Section basic properties solution Yang Yang} 
we review several known properties of the solution $\veps(\la)$ to  \eqref{ecriture eqn YY} 
and then establish some uniform in $T$ bounds on the location of its zeroes. 
Then, several properties of integral operators that appear handy in the course of the proof of Theorem 
\ref{Theorem existence of AE for YangYang energy}
are given in Section \ref{Section Prop of Lin Int Ops}. 
Finally,  in Section \ref{Section low-tempe}, we prove Theorem \ref{Theorem existence of AE for YangYang energy}.

\section{Several properties of the solution to the Yang-Yang equation}
\label{Section basic properties solution Yang Yang}

We start this section by recalling, for the reader's convenience and self-sufficiency of the paper, Yang-Yang's proof for the 
existence of solutions to  \eqref{ecriture eqn YY}. We also provide a simple argument 
so as to justify the uniqueness of solutions. 
We then push further the analysis of the objects involved in the proof of the existence of solutions
what allows us to provide $T$-independent bounds for the zeroes of $\veps(\la)$. These
bounds will play a role in the next section.

\begin{theorem} \cite{Yang-YangNLSEThermodynamics}
\label{Theorem Yang-Yang existence solutions}
The Yang-Yang equation admits a unique solution $\veps(\la)$ such that $[\veps(\la)-\la^2 ] \in \msc{L}^{\infty}(\R)$ and is entire.
This solution is even, strictly increasing on $\intoo{0}{+\infty}$
and satisfies to the bounds $\veps^{\prime}(\la) > \la $ for any $\la \in \intoo{0}{+\infty}$. 
Lastly, for $h>0$, $\veps$ admits a unique zero $\wh{q}$ on $\intoo{0}{+\infty}$. 

\end{theorem}

\Proof

The Yang-Yang equation can be recast as the fixed point equation $\mc{L}[v](\la) =  v(\la)$
for the functional $\mc{L} : L^{\infty}(\R) \tend L^{\infty}(\R)$ given by 
\beq
\mc{L}[f](\la) \; =  \; - \;  h  \; - \; \f{T}{2\pi} \Int{ \R }{} K(\la-\mu) \ln \big[ 1+ \ex{-\f{\mu^2  + f(\mu)}{T}}\big] 
\dd \mu  \;. 
\enq
Indeed, let $v$ correspond to a fixed point of $\mc{L}$, then it is readily seen that $\veps(\la) = v(\la) + \la^2 $
solves the Yang-Yang equation. The matter is that $\mc{L}$ has at most one fixed point. Indeed, 
given $f,g \in L^{\infty}(\R)$ 
\beq
\Big| \mc{L}[f](\la) - \mc{L}[g](\la)\Big| \; = \; \bigg|  \Int{\R}{}  \Int{0}{1} 
		\f{K(\la - \mu)  [g(\mu)-f(\mu)] }{  1+  \ex{ \f{\mu^2}{T} + \f{(1-s)}{T}g(\mu) + \f{s}{T}f(\mu) }  }  \dd s \f{ \dd \mu }{ 2\pi }
				\bigg|  \; \leq  \;  \f{ \norm{f-g}_{L^{\infty}(\R)}  }
{ 1 + \exp\big\{ -\f{1}{T}\e{max}( \norm{f}_{\infty}, \norm{g}_{\infty} ) \big\} } 
\enq
As a consequence, $\norm{\mc{L}[f] -\mc{L}[g] }_{L^{\infty}(\R)} < \norm{f -g }_{L^{\infty}(\R)}  $, 
so that the sharpness of the inequality ensures the uniqueness of a fixed point in $L^{\infty}(\R)$. 
Note that two distinct solutions to the Yang-Yang equation would give rise to two distinct fixed points of $\mc{L}$.
Thus, there is at most one solution to the Yang-Yang equation as well. 

We now build a sequence converging to the fixed point of $\mc{L}$. Following Yang and Yang,
the sequence $v_n(\la)$ is defined as $v_0(\la)= 0$, $v_1(\la)=-h$ and 
$v_{n+1}(\la)=\mc{L}[v_n](\la)$ for $n \geq 1$. The sequence $v_n(\la)$ is such that 
\begin{itemize}
\item $v_n(\la)$ is a decreasing sequence in $L^{\infty}(\R)$ of even functions of $\la$;
\item $v_n(\la)$ is smooth and $v^{\prime}_n(\la) \geq  0$ for any $\la \in \intoo{0}{+\infty}$ .
\end{itemize}
These properties are readily seen to hold for $n=1$ 
%
%
%
%
%
%
%
%
Now, assume that these properties hold for some $n\in \mathbb{N}^*$. 
Since $v_n \in L^{\infty}(\R)$, one has that $\mc{L}[v_n](\la)$ is also in $L^{\infty}(\R)$. 
Derivation under the integral theorems guarantee that $\mc{L}[v_n](\la)$ is smooth. 
Since the integrand is negative, it is clear that $v_2-v_1 <0$. Else, for $n \geq 2$ one has 
\beq
v_{n+1}(\la) \; - \;  v_{n}(\la)  
%
%
%
\; = \;  \Int{ \R }{} K(\la-\mu)  \Bigg\{ \Int{ v_{n-1}(\mu) }{ v_n(\mu) } \f{ \dd s }{ 1+\ex{\f{s + \mu^2}{T}} }  \Bigg\} \cdot 
				 \f{\dd \mu}{2\pi} <0 \;, 
\nonumber
\enq
and, using that $v^{\prime}_n(\la)$  is odd, one has for $\la \in \intoo{0}{+\infty}$ 
\beq
v_{n+1}^{\prime}(\la)  \; = \;  \Int{ \R }{}  \f{ K(\la-\mu) [ 2\mu + v_n^{\prime}(\mu) ] }
				{ 1 + \ex{\f{ v_n(\mu) \, + \, \mu^2 }{T} } } 
\cdot  \f{ \dd \mu }{ 2\pi } 
\; = \;  \Int{ 0}{ +\infty } \big[ K(\la-\mu) - K(\la+\mu) \big]  \cdot 
\f{  2\mu + v_n^{\prime}(\mu)  }
				{ 1 + \ex{\f{ v_n(\mu) \, + \, \mu^2 }{T} } } 
\cdot  \f{ \dd \mu }{ 2\pi } >0 \;, 
\enq
 since, on the one hand $v_n^{\prime}(\mu)\geq 0$ on $\intff{ 0 }{ +\infty }$ and on the other hand
$K(\la-\mu) - K(\la+\mu) >0$ for $(\la,\mu) \in (\R_+^*)^2$.
\vspace{2mm}

%
%
%
%
%
%
%
%
%
%
%

As $\la \mapsto v_n(\la)$ is increasing, it follows that $v_n(\la) \geq  v_n(0)$, so that 
\beq
v_{n+1}(0)  \; \geq \; L( v_n(0)  ) \qquad \e{with} \qquad 
L(x) \; = \;  - h \; -\;  T \Int{\R}{}\hspace{-1mm} K(\mu) \ln\Big[ 1+ \ex{-\f{\mu^2}{T} - \f{x}{T} } \Big] \cdot \f{ \dd \mu  }{2\pi} \;. 
\label{definition fonction L}
\enq
The function 
\beq
L(x)-x \; =  \; - h  \; -\;  T \Int{\R}{} K(\mu) \ln\Big[ \ex{ \f{x}{T} }+ \ex{-\f{\mu^2}{T}   } \Big] \cdot \f{ \dd \mu  }{2\pi} \;.
\enq
is strictly decreasing. The dominated convergence theorem ensures that 
\beq
\lim_{x\tend +\infty} L(x) = -h  \qquad \e{meaning} \; \e{that} \;  \qquad \lim_{x\tend +\infty} \big[ L(x) - x \big] = -\infty \;. 
\enq
Moreover given any decreasing sequence $x_n$ such that  $x_n \tend -\infty$, one has that the increasing sequence of functions 
$-\ln \Big[ \ex{ \f{x_n}{T} } \, + \, \ex{ -\f{\mu^2}{T} } \Big] K(\mu)$
%
%
%
%
%
%
%
%
%
%
%
is bounded from below by the integrable function $-\ln \Big[ \ex{ \f{x_0}{T} }  \, + \, \ex{-\f{\mu^2}{T} }  \Big]K(\mu)$. 
Hence, by the monotone convergence theorem
\beq
\lim_{n\tend +\infty} \Int{\R}{}  -K(\mu) \ln \Big[ \ex{\f{x_n}{T} } \, + \, \ex{-\f{\mu^2}{T} } \Big]  \cdot  \f{ \dd \mu }{2\pi} \; = \; 
\Int{\R}{}  \f{ K(\mu)\mu^2 }{2\pi T}  \cdot \dd \mu \; = \; +\infty \;. 
\enq
Thus $\lim_{x\tend -\infty} \big[ L(x)-x \big] = +\infty$. Since $L(x)-x$ is continuous on $\R$ and strictly decreasing, 
it admits a unique zero $-z_h$ on $\R$. Moreover, since $L(0)<-h$ and $L(x)-x$ is strictly decreasing,  
it follows that $z_h>0$ for $h>0$. 
One can also check that the function $L$ is strictly increasing on $\R$. 
%
%
%
%
%
%
%
%
Thus,
\beq
\lim_{x \tend +\infty} L(x) \;  = \; -h \quad  \Rightarrow \quad -h > L(x) \quad \e{for}\;\e{any} \;  x \in \R\;.  \nonumber
\enq
As a consequence, one has
$-h > L(-z_h)= - z_h$. Then, a straightforward induction shows that 
$v_n(0)> -z_h$. Since $v_n(\la) \geq v_n( 0 ) > - z_h$, one has that $v_n(\la)$ is a decreasing sequence
of measurable functions that is bounded from below. Thus $v_n(\la) \tend v(\la)$ pointwise to a measurable function 
$v(\la)$. Taking the pointwise limits in the relations
\beq
v_{n}(\la) \geq - z_h \quad \e{and} \quad  v_n(\la)-v_{n}(-\la)=0 
\enq
 ensures that $v(\la)$ satisfies to the same properties. As the sequence $v_n$ is decreasing, $0=v_0(\la)\geq v(\la) $, 
and thus $v \in L^{\infty}(\R)$.

It is then enough to apply the monotone convergence theorem so as to get that $v(\la)$ corresponds to the fixed point of 
$\mc{L}$ in $L^{\infty}(\R)$. Indeed, for any $\la$, one has that 
\beq
\mu \mapsto h_n(\mu) \; = \; K(\la-\mu) \ln \big[ 1 \,+ \, \ex{- \f{ v_{n}(\mu) + \mu^2 }{ T } } \big]
\enq
is an increasing sequence of positive measurable functions which, due to $v_{n}(\la) \geq  v_n(0)  \geq - z_h$, 
is bounded from above by a $L^{1}(\R)$ function
\beq
h_n(\mu) \leq K(0) \ln \big[ 1  \, +  \,  \ex{- \f{ \mu^2  - z_h }{ T } }   \big] \;. 
\enq
Since $h_n$ converges pointwise, one has that 
\beq
 \lim_{n\tend +\infty} \Int{\R}{} \! h_n(\mu)  \cdot \f{ \dd \mu }{ 2\pi } \; = \; 
\Int{\R}{} K(\la-\mu) \ln \big[ 1 \,+ \, \ex{- \f{ v(\mu) + \mu^2 }{ T } } \big]  \cdot \f{\dd \mu }{ 2\pi } \;. 
\enq
Taking the pointwise limit in $v_{n+1}(\la) \; = \;  \mc{L}[v_n](\la)$ shows that $v $ is 
indeed the fixed point of $\mc{L}$. 

Finally, it follows from the integral representation given by 
$v(\la) = \mc{L}[v](\la)$ that $v$ is holomorphic in the strip $\mc{S}_{c}$. By using this information
and deforming the integration contour $\R$ in $\mc{L}[v](\la)$, it is readily seen that $v$ is holomorphic in $\mc{S}_{2c}$,
its analytic continuation being given by
\beq
v(\la) \; = \; \; - \; \f{T}{2\pi} \Int{ \R \pm i c}{} K(\la-\mu) \ln \big[ 1+ \ex{-\f{\mu^2  + f(\mu)}{T}}\big] 
\dd \mu  \qquad \e{for} \qquad 2c \, >  \, \pm \Im(\la) \, > \,  0\;.
\label{equation continuation analytique fct v}
\enq
By immediate induction based on deforming the contour in \eqref{equation continuation analytique fct v} further, 
one gets that $v$ is entire. A derivation under the integral sign ensures that $v^{\prime}(\la) >0$ 
for $\la \in \intoo{0}{+\infty}$. 
Hence, setting $\veps(\la) = \la^2 + v(\la)$ we get the sought unique solution to the Yang-Yang equation. 
Moreover, we immediately have that the latter satisfies $\veps^{\prime}(\la)>0$ on $\R^+$, $\veps(0)< - h $ and 
$\veps(\la) \geq -z_h + \la^2 $. Thus $\lim_{\la \tend +\infty}\veps(\la) = +\infty$, so that $\veps$
admits a unique zero $\wh{q}$ on $\R^+$. Hence, by eveness, it has thus two zeroes $\pm \wh{q}$ on $\R$. 

\qed

\begin{prop}
\label{Proposition bounds zero veps}
Let $h>0$, then there exists $T_0$ such that for any $T\in \intff{0}{T_0}$ the unique zero $\wh{q}$ of $\veps$
as defined in Theorem \ref{Theorem Yang-Yang existence solutions} satisfies $\sqrt{h} < \wh{q} < 2 \sqrt{w}$, where 
$w \in \intoo{0}{+\infty}$ corresponds to the unique zero of 
\beq
\om(x) \; = \; x  \, - \, h \, - \, \f{ 2 (x +c^2)}{\pi}  \arctan \Big(  \f{\sqrt{x}}{c} \Big)   \;  + \;   \f{ 2 c }{\pi } \sqrt{x} \;. 
\enq

\end{prop}

\Proof
Evaluating the Yang-Yang equation at $\la=\wh{q}$ leads to 
\beq
\wh{q}^{\,2} \; = \;  h  \; + \;  \f{T}{2\pi} \Int{ \R }{} K(\,\wh{q}-\mu) \ln \big[ 1+ \ex{-\f{\veps(\mu)}{T}}\big] \cdot \dd \mu  \; . 
\enq
The second term is positive meaning that $ \wh{q} \,  \geq  \sqrt{h}$, this uniformly in $T\in \intfo{0}{+\infty}$.

It thus remains to obtain the upper bound. It was established in the course of the proof 
of Theorem \ref{Theorem Yang-Yang existence solutions}
that $\veps(\la) \geq \la^2  - z_h$ where $z_h$ solves $-z_h = L(-z_h)$ with $L$ defined in \eqref{definition fonction L}. 
Thus $\wh{q} \leq \sqrt{z_h}$ and the upper bound will follow from an estimate on $z_h$. 
For this purpose, we recast the defining equation for $z_h$ in the form
\beq
z_h  \; = \;  h + \Int{0}{\sqrt{z_h}} (z_h-\mu^2) K(\mu) \cdot \f{ \dd \mu }{ \pi } \; + \; V_h(T) \qquad \e{where} \qquad 
V_h(T)  \; = \;  \f{T}{\pi} \Int{0}{+\infty} K(\mu) \ln \big[ 1 + \ex{ - \f{ |\mu^2-z_h | }{ T } }  \big] \cdot \dd \mu
\label{ecriture equation pour zh}
\enq

It is readily seen that $V_h(T) \leq T \ln 2$. Explicit integration of the second term in \eqref{ecriture equation pour zh} 
shows that $z_h$ solves the equation $\om(z_h) = V_h(T)$. 
The function $\om$ is smooth on $\intoo{0}{+\infty}$ and strictly increasing 
\beq
\om^{\prime}(x) \; =\; 1 - \f{2}{\pi} \arctan\Big( \f{\sqrt{x} }{ c } \Big) >0. 
\enq
Since $\om(0)=-h$ and $\om ( x ) \sim \tf{4 c \sqrt{x} }{\pi}  $ as $x\tend +\infty$, 
it follows that $\om$ admits a unique zero $w$ on $\intoo{0}{+\infty}$. 
Moreover $\om$ is a $\mc{C}^{\infty}$-differomorphism from $\intoo{0}{+\infty}$ onto $\intoo{-h}{+\infty}$. 
Since $V_h(T)>0$, one has that $V_h(T) \in\intoo{-h}{+\infty}$ so that $z_h = \om^{-1}\big(V_h(T) \big) $.
Then, the Taylor expansion of $\om^{-1}(t)$ to the first order around $t=0$ 
and the estimates for $V_h(T)$ ensure that $z_h = w + \e{O}(T)$. 
In particular, as $\om^{-1}$ is strictly increasing, it follows that there exists $T_0$ such that $w \leq z_h \leq 4 w$ 
on $\intff{0}{T_0}$. As a consequence, $\wh{q} \leq 2 \sqrt{w}$ uniformly in $T \in \intff{0}{T_0}$. \qed


\section{Some properties of a linear integral equations driven by the Lieb kernel}
\label{Section Prop of Lin Int Ops}

\subsection{Invertibility of the Lieb operator}

\begin{lemme}

Let $K(\la)$ be the Lieb kernel defined in \eqref{ecriture eqn YY} and $f\in L^{1}(\intff{-\a}{\a})$, then 
\beq
 \Big| \Int{-\a}{\a} f(\mu) K(\la-\mu) \dd \mu \dd \la \Big| \; \leq  \; 
 \Int{-\a}{\a}  |f(\mu)| \, \dd \mu \cdot \Int{-\a }{\a } \! K(\la) \,  \dd \la  \;,
\label{equation majorant integrale convolution avec K}
\enq
\end{lemme}

\Proof

One has 
\beq
 \Big| \Int{-\a}{\a}\!  f(\mu) K(\la-\mu) \dd \mu \dd \la \Big| \; \leq  \; 
\Int{-\a}{\a} \! \dd \mu  \bigg\{ |f(\mu)| \Int{-\a-\mu }{\a -\mu }\!  K(\la) \,  \dd \la  \bigg\} \;.
\enq
The integral involving the kernel $K$ can be decomposed as
\beq
\Int{-\a-\mu }{\a -\mu } \!\!\!   K(\la)  \cdot  \dd \la   \; =  \;  \Int{-\a }{\a } \!\!   K(\la) \cdot  \dd \la  
\; + \; \Int{-\a -\mu }{- \a } \!\!\!   K(\la) \cdot  \dd \la   \; - \;  \Int{\a-\mu}{\a} \!\!\!   K(\la) \cdot  \dd \la \;. 
\label{equation pour majorer integrale K}
\enq
Yet, for any $\mu \in \intff{-\a}{\a}$, one has that 
\beq
\Int{-\a - \mu }{- \a }\! K(\la) \cdot  \dd \la    \;  - \;  \Int{\a - \mu}{\a}\! K(\la) \cdot  \dd \la \; = \; 
\Int{ \a }{\a + \mu } \! [ K(\la) - K(\la-\mu) ] \cdot  \dd \la     \; \leq \;  0 \;. 
\enq
This is evident for $\a \geq \mu\geq 0$, since then $[ K(\la) - K(\la - \mu  ) ] \leq 0$ for any $\la \geq 0$. 
The case when $- \a \leq \mu\leq 0$ follows since the orientation of the integral produces a sign and 
$[ K(\la) - K(\la - \mu  ) ] = [ K(\la) - K(\la + \abs{\mu}  ) ]\geq 0$ for $\la\geq 0$. 
As, the last two terms in \eqref{equation pour majorer integrale K}
produce negative contributions, one gets 
\beq
\Int{-\a-\mu }{\a -\mu } \! K(\la) \,   \dd \la  \; \leq  \;  \Int{-\a }{\a } \! K(\la) \,  \dd \la   \;.  
\enq
\qed

\begin{lemme}
\label{Lemme existence dvpm series det et resolvent}
Let $\a \in \intfo{0}{+\infty}$ . The integral operator $\tf{K}{(2\pi)}$ acting on $L^{2}(\intff{-\a}{\a})$ with 
the integral kernel $K(\la-\mu)$ given in \eqref{ecriture eqn YY} is a compact  trace class operator. 
The logarithm of the Fredholm determinant  of $I-\tf{K}{(2\pi)}$ is well defined and given by the convergent series
\beq
\ln \det_{\intff{ -\a }{ \a } }\Big[ I-\f{K}{2\pi} \Big]  = - \f{ 2\a K(0)  }{ 2\pi }
\; - \; \sul{n=2}{+\infty} \f{1}{n} \Int{-\a}{\a}  \pl{\ell=1}{n} K(\tau_{\ell}-\tau_{\ell+1})  \cdot \f{ \dd^n \tau }{ (2\pi)^n} \;, 
\label{ecriture log det I+K}
\enq
where in each $n$-fold integral periodic boundary condition $\tau_{n+1} = \tau_1$ are imposed. 
Finally, the operator $I-\tf{K}{(2\pi)}$ is invertible with inverse  $I + \tf{ R^{(\a)}}{ (2\pi) }  $.
The kernel $R^{(\a)}(\la, \mu)$ of the resolvent operator $R^{(\a)}$ can be expressed in terms of the below 
series of multiple integrals
\beq
R^{(\a)}(\la, \mu) \; = \; 
K(\la-\mu) \; + \; \sul{n = 1}{+\infty} \Int{ -\a }{ \a } K(\la-\tau_1) \cdot \pl{\ell=1}{n-1} K(\tau_{\ell}-\tau_{\ell+1})
 \cdot K(\tau_{n}-\mu) \cdot \f{ \dd^n \tau }{ (2\pi)^n }\;. 
\label{definition resolvent}
\enq
Given any $0<a<\tf{c}{2}$, the series representation \eqref{definition resolvent} is convergent in respect to the sup norm 
on $\mc{S}_a \times \mc{S}_a$ (with $\mc{S}_a$ as in \eqref{definition strip width a}).
Moreover, for any $(\la,\mu) \in \mc{S}_{a} \times \mc{S}_{a}$ with $0<a<\tf{c}{2}$, one has the estimates 
\beq
| R^{(\a)}(\la, \mu) | \; \leq  \;  \norm{ K }_{L^{\infty}(\mc{S}_{2a})} 
\cdot\bigg\{ 1\; + \;   \f{ C_K }{ 1 \; - \;  \norm{ \tf{K}{(2\pi)} }_{L^{1}( \intff{-\a}{\a} ) }  }  \bigg\} \;\qquad \e{where} \qquad
C_K \; = \; \e{sup}_{\la \in \mc{S}_a} \Int{ -\a }{ \a } |K(\la-\tau)|  \cdot \f{ \dd \tau }{2\pi }
\label{ecriture bornes resolvent sur interval general}
\enq
and the resolvent kernel is an analytic function of $(\la,\mu)$ 
on the strip $\mc{S}_{\tf{c}{2}} \times \mc{S}_{\tf{c}{2}}$  

Finally, the function $(\a, \la, \mu) \mapsto R^{(\a)}(\la,\mu)$ is $\mc{C}^{1}$ on $\R^3$. 

\end{lemme}

\Proof 

The estimates for any of the $n$-fold multiple integrals occurring in 
\eqref{ecriture log det I+K}-\eqref{definition resolvent} can be obtained by using the majoration
\eqref{equation majorant integrale convolution avec K}. 
We treat the case of the $n$-fold cyclic 
integral occurring in the series expansion for \newline $\ln\det[I-\tf{K}{(2\pi)}]$. For $n \geq 2 $
\bem
\Int{-\a}{\a}  \pl{\ell = 1}{n} K(\tau_{\ell} - \tau_{\ell +1 } )  \cdot  \f{ \dd^n \tau }{ (2\pi)^n } \;  \leq  \; 
K(0) \Int{-\a}{ \a} K(\tau_1-\tau_2) \dots K(\tau_{n-1}-\tau_n) 
\cdot \f{ \dd^n \tau }{ (2\pi)^n }  \\ 
\; \leq  \; 
\f{ K(0) }{ (2\pi)^n } \cdot   \norm{K}_{ L^1(\intff{-\a}{\a}) } \Int{-\a}{ \a} K(\tau_1-\tau_2) \dots K(\tau_{n-2}-\tau_{n-1}) \cdot \dd^{n-1} \tau
\; \leq  \; \f{ 2\a  K(0) }{ (2\pi)^n } \norm{K}_{L^1(\intff{-\a}{\a})}^{n-1}  \;. 
\end{multline}
It is clear that these bounds also hold for $n=1$. Further, since
\beq
\Int{\R}{} \! K(\la) \,  \f{ \dd \la }{ 2\pi }\;  = \;  1 \; \qquad \e{it} \; \e{follows} \; \e{that} 
\qquad  \norm{  K   }_{ L^1(\intff{-\a}{\a}) } \!\! < \; 2\pi \;. 
\enq
As a consequence, the series representation for $\ln \ddet{}{I-\tf{K}{(2\pi)}}$ is convergent. 
The same types of estimates also ensure the uniform convergence
in respect to the sup norm topology on compact subsets of $\R^2$ of the series \eqref{definition resolvent} defining the resolvent. 

Lastly, the fact that $ (\a,\la,\mu) \mapsto R^{(\a)}(\la,\mu) \in \mc{C}^{1}(\R^3)$  can be readily inferred from the 
Fredholm series representation for $R^{(\a)}(\la,\mu)$, $\ddet{ \intff{-\a}{\a} }{ I-\tf{K}{(2\pi)}}$
and the fact that $\ddet{ \intff{-\a}{\a} }{ I-\tf{K}{(2\pi)}}\not=0$ for any given $\a \in \intfo{0}{+\infty}$. \qed

\subsection{Positivity property of Lieb-like kernels}

\begin{lemme}
\label{Lemme positivite difference resolvant}
Let $I- O$ be a trace class integral operator on $L^{2}(\intff{-a}{ a })$ with $0 < a < +\infty$
whose integral kernel is continuous and satisfies 
\beq
O(\la,\mu) \, - \,  O(\la,-\mu) \; \geq  \; 0 \qquad \e{and} \qquad O(\la,\mu)=O(-\la,-\mu) \qquad \e{for}\;  any \qquad \mu,\;  
\la \in \intoo{0}{a} \;. 
\enq
Assume that the series representation 
\beq
R_O(\la,\mu) = O(\la,\mu) \; + \;  \sul{n \geq 1}{} \; \Int{-a}{a} \! O(\la,\tau_1) \cdot  
\pl{\ell = 1}{n-1} \Big\{ O(\tau_{\ell},\tau_{\ell + 1}) \Big\} \cdot 
 O(\tau_{n},\mu)   \cdot \dd^n \tau
\enq
for the resolvent 
kernel $R_O(\la,\mu)$ of the inverse operator $I+R_O$ to $I-O$  is convergent in $L^{\infty}([-a;a]^2)$. 
Then, the resolvent kernel $R_O(\la,\mu)$  satisfies 
\beq
R_O(\la,\mu)- R_O(\la,-\mu) \; \geq \; 0 \qquad  for  \;  any \qquad \mu \, , \la  \; \in \; \intoo{0}{a} \;. 
\enq

\end{lemme}

\Proof

We consider the series representation for $R_O(\la,\mu)- R_O(\la,-\mu)$ and show that every term in this series is positive. 
Due to the hypothesis of the lemma regarding to the kernel $O(\la,\mu)$, it is enough to show that 
\bem
\Int{-a }{ a } 
O(\la,\tau_1)\pl{ \ell  =1}{n-1}O(\tau_{ \ell },\tau_{ \ell  +1}) \cdot \big[  O(\tau_{n},\mu)  -  O(\tau_{n},-\mu) \big] \cdot \dd^n \tau  \\
\; = \;  
\Int{0}{a} \big[  O(\la,\tau_1)  -  O(\la,-\tau_1) \big]
 \pl{\ell  =1}{n-1}\big[  O(\tau_{\ell  }, \tau_{\ell  +1})  -  O(\tau_{\ell},-\tau_{\ell + 1}) \big]
 \cdot \big[  O(\tau_{n},\mu)  -  O(\tau_{n},-\mu) \big]  \cdot \dd^n \tau  \;. 
\label{equation positivite terme n diff resolvent}
\end{multline}
For $n=1$, one has that 
\beq
\Int{-a}{a} \! O(\la,\tau) \big[  O(\tau,\mu)  -  O(\tau,-\mu) \big]  \cdot \dd \tau = 
\Int{0}{a}  \! O(\la,\tau) \big[  O(\tau,\mu)  -  O(\tau,-\mu) \big] \cdot  \dd \tau  \; + \; 
\Int{0}{a} \! O(\la,-\tau) \big[  O(-\tau,\mu)  -  O(-\tau,-\mu) \big]  \, \cdot \, \dd \tau  
\enq
Above, we have split the integration interval, into the parts $\intff{0}{a}$ and $\intff{-a}{0}$
and have changed the variables $\tau \mapsto -\tau^{\prime}$  in the integral over the segment $\intff{-a}{0}$. 
It then remains to use the reflection invariance of the kernel $O(-\tau,-\mu) = O(\tau,\mu)$ so as to get the representation for
$n=1$. Assume that \eqref{equation positivite terme n diff resolvent} holds for some $n$. The
ultimate integration over $\tau_{n+1}$ occurring in the $n+1$-fold integration can be recast in much the same way as for $n=1$:
\beq
\Int{-a}{a} \! O(\tau_n,\tau_{n+1}) \big[  O(\tau_{n+1},\mu) \;  -  \;   O(\tau_{n+1},-\mu) \big]  \cdot \dd \tau_{n+1} = 
\Int{0}{a}\!  \big[  O(\tau_n,\tau_{n+1}) \;  - \; O(\tau_n,-\tau_{n+1}) \big] \cdot  
\big[  O(\tau_{n+1},\mu)  -  O(\tau_{n+1},-\mu) \big]  \cdot \dd \tau_{n+1} \;. 
\nonumber
\enq
Once this operation is performed, the integration with respect to $\tau_1, \dots, \tau_n$ is of the form
taken care of by the induction hypothesis. The result is proven for $n+1$. 

It then follows that 
\bem
R_O(\la,\mu) \,- \, R_O(\la,-\mu) =  O(\la,\mu) \,- \, O(\la,-\mu)  \\
\; + \;  \sul{n \geq 1}{} \Int{0}{a}  \!  \big[  O(\la, \tau_{1})  -  O(\la,-\tau_{1}) \big] \cdot 
\pl{\ell =1}{n-1}\big[  O(\tau_{\ell  }, \tau_{\ell +1})  -  O(\tau_{\ell },-\tau_{\ell +1}) \big]  
\cdot \big[  O(\tau_{n}, \mu)  -  O(\tau_{n},-\mu) \big]   \cdot \dd^n \tau 
\;  . 
\end{multline}
 Since every term in this series is positive, the claim follows. \qed

\subsection{Existence and uniqueness of the dressed energy}

In the Bethe Ansatz literature, the dressed energy refers to the unique couple $(\veps_0(\la\mid q), q)$ solving the conditions
\beq
\veps_0(\la \mid q ) \; - \;  \Int{ - q }{  q } K(\la-\mu) \veps_0(\mu \mid q ) \cdot \f{ \dd \mu }{2\pi} =  \la^2 - h  \qquad \e{where} 
\; q \;\e{is} \; \e{such} \; \e{that} \qquad 
\veps_0(q \mid q) = 0 \;. 
\label{ecriture equation veps GS et definition zero q}
\enq

Lemma \ref{Lemme existence dvpm series det et resolvent} ensures the existence and uniqueness of solutions
$g  \in L^2(\intff{-\a}{\a}) $ to all integral equations of the type 
\beq
\big(I-\tf{K}{(2\pi)} \big)\cdot g \; =\;  f \;, \qquad  \e{for} \; \e{any} \quad  f \in L^2(\intff{-\a}{\a}) \;. 
\enq
However, the unique solvability of \eqref{ecriture equation veps GS et definition zero q} is a more involved issue
in as much as the endpoint of integration $q$ is also subject to a non-linear equation. 
In fact the statement about the existence of a solution to the system \eqref{ecriture equation veps GS et definition zero q}
has been made at several places in the litterature 
\cite{BogoliubiovIzerginKorepinBookCorrFctAndABA,BortzLowTExpansionAtNonZeroMagField,EsslerFrahmGohmanKlumperKorepinOneDimensionalHubbardModel,
IidaWadaitYangYangEqnAnalysisAtT=0,KlumperScheerenSomeLowTComputationsXXX} 
altough, to the best of the author's knowledge, no proof has ever been given.  
In the proposition below, we prove the unique solvability of the system \eqref{ecriture equation veps GS et definition zero q}.

\begin{prop}
\label{Proposition preuve existence q pour epsilon 0}

Let $\veps_0(\la\mid \a)$ denote the unique solution to the linear integral equation
\beq
\veps_0(\la\mid \a) \; - \;  \Int{ - \a }{  \a } K(\la-\mu) \, \veps_0(\mu\mid \a) \cdot \f{ \dd \mu }{2\pi} =  \la^2 - h   \;. 
\label{equation int veps alpha}
\enq
Given any $\a$, the function $\la\mapsto \veps_0(\la\mid \a)$ is analytic on the strip 
$ \mc{S}_{ \tf{c}{2} }$.  

Also, there exists a unique $q \in \intoo{0}{+\infty}$ such that $\veps_0(q \mid q)=0$. 
Moreover, there exists open neighbourhoods $O_q$ of $q$ and $V_0$ of $0$ in $\R$ such that 
the function $\a \mapsto \veps_0(\a \mid \a)$ is a $\mc{C}^{\infty}$-diffeomorphism from $O_q$ onto $V_0$.

\end{prop}

\Proof 

Let $R^{(\a)}(\la,\mu)$ be the resolvent kernel  associated with  $I-\tf{K}{(2\pi)}$ understood as acting on $L^{2}(\intff{-\a}{\a})$. 
By lemma \ref{Lemme existence dvpm series det et resolvent}, it is an analytic function of $(\la,\mu) \in \mc{S}_{\tf{c}{2}}\times
\mc{S}_{\tf{c}{2}}  $. Since
\beq
\veps_0(\la \mid \a) \; = \; \la^2 - h \; +  \; \Int{-\a}{\a} R^{(\a)}(\la,\mu) \cdot (\mu^2 -h ) \cdot \f{ \dd \mu }{ 2\pi}  \;, 
\label{ecriture rep veps la alpha en terms resolvent}
\enq
the statement about the analyticity of $\la \mapsto \veps_0(\la \mid \a)$ on $\mc{S}_{\tf{c}{2}}$ follows. 
Moreover, since $(\a,\la,\mu) \mapsto R^{(\a)}(\la,\mu)$ is $\mc{C}^{1}$ and the integration in 
\eqref{ecriture rep veps la alpha en terms resolvent} goes through the compact interval $\intff{-\a}{\a}$, 
it follows by differentiation under the integral theorems that $\veps_0(\la\mid \a)$ is continuously 
differentiable in respect to $\a$.

In order to prove the existence and uniqueness of $q$, we study the variations of the function $\a \mapsto \veps_0(\a \mid \a)$. 
For this, we first obtain a representation for its derivative:
\beq
\f{\dd}{ \dd \a} \veps_0(\a \mid \a) \;  =  \; { \f{\Dp{}  }{ \Dp{} \la} \veps_0(\la \mid \a) }_{  \mid \la=\a}
\; + \; { \f{\Dp{}  }{ \Dp{} \a_0} \veps_0(\la \mid \a_0) }_{\mid \a_0=\a}\;. 
\enq
It is readily seen through integration by parts that the partial derivatives satisfy the integral equations
\beqa
\Big[ \Big( I - \f{K}{2\pi} \Big)\cdot \Dp{\a}\veps_0 \Big] (\la \mid \a) &=& 
\f{\veps_0(\a,\a)}{2\pi} \Big[K(\la-\a) \; + \;K(\la+\a)  \Big]  \; \\  
\Big[ \Big( I - \f{K}{2\pi} \Big)\cdot \Dp{\la}\veps_0 \Big] (\la \mid \a) &= &
2 \la - \f{\veps_0(\a,\a)}{2\pi} \Big[K(\la-\a) \; - \;K(\la+\a)  \Big] \;. 
\eeqa
These are readily solved in terms of the resolvent kernel 
\beqa
\f{ \Dp{} }{ \Dp{} \a } \veps_0  (\la,\a) &=& \f{\veps_0(\a \mid \a)}{2\pi} \Big[R^{(\a)}(\la,\a) \; + \;R^{(\a)}(\la,-\a)  \Big]  
\label{ecriture derivee eps0 alpha} \\
 \f{ \Dp{} }{ \Dp{} \la } \veps_0  (\la,\a) &=& - \f{\veps_0(\a \mid \a)}{2\pi} \Big[R^{(\a)}(\la,\a) \; - \;R^{(\a)}(\la,-\a)  \Big] \; + \; 
2\la  \; + \;  \Int{-\a}{\a} R^{(\a)}(\la,\mu) \cdot \f{2\mu}{2\pi} \cdot \dd \mu \;. 
\label{ecriture derivee eps0 lambda} 
\eeqa 

As a consequence, 
\beq
\f{ \dd }{ \dd \a} \veps_0(\a \mid \a) = \f{\veps_0(\a \mid \a) }{ \pi } R^{(\a)}(\a,-\a) \; + \;  2\a  \; + \; 
\Int{0}{\a} \big[R^{(\a)}(\la,\mu) - R^{(\a)}(\la,-\mu) \big] \cdot \mu  \cdot  \f{\dd \mu}{\pi} 
\label{ecriture forme explicite derivee eps alpha alpha}
\enq
It follows from lemma \ref{Lemme positivite difference resolvant} that the integrand of the integral appearing 
in the \textit{rhs} is strictly positive on $\intoo{0}{+\infty}$. It follows from \eqref{definition resolvent}
that $R^{(\a)}(\la,\mu) >0$. Thus  $\f{ \dd }{ \dd \a} \veps_0(\a \mid \a) >  0$ as soon as $\veps_{0}(\a\mid \a) \geq 0$. 

Suppose that there exist a $q>0$  such that $\veps_0(q,q)=0$. It then follows from 
\eqref{ecriture forme explicite derivee eps alpha alpha} that $\a \mapsto \veps_0(\a \mid \a)$ is strictly increasing 
in some open neighbourhood $\intoo{q-\eta}{ q + M }$ of $q$ with $\eta>0, M>0$. Let $M$ correspond to the largest real 
with this property and suppose that $M \not= +\infty$. One has that $\veps_0(\a\mid \a)$ is strictly increasing
on $\intoo{q}{q+M}$. Hence, by continuity $\veps_0(M+q \mid M+q) >0$. Thus, one has that  
\beq
{  \f{ \dd }{ \dd \a} \veps_0(\a \mid \a) }_{\mid \a =q+M } >0 \;, 
\enq
contradicting the definition of $M$. It thus follows that $ \veps_0(\a \mid \a) >0$ 
on $\intoo{q}{+\infty}$. This reasoning shows that $\a \mapsto \veps_0(\a \mid \a)$  has at most one zero on 
$\intff{0}{+\infty}$. It thus remains to show that it has at least one. It follows from 
\eqref{ecriture equation veps GS et definition zero q} that $\veps_0(0 \mid 0)=-h<0$.
It thus remains to show that $\veps_0(\a\mid \a) >0$  for $\a$ large enough. 
For this, recast \eqref{ecriture rep veps la alpha en terms resolvent} as 
\beq
\veps_0(\a \mid \a)  = \a^2 - h \; + \; 
 \Int{ \sqrt{h} }{ \a } \big[ R^{(\a)}(\a,\mu) + R^{(\a)}(\a,-\mu) \big] (\mu^2 -h) \cdot \f{ \dd \mu }{2\pi} 
\; - \; \Int{0}{ \sqrt{h} } \big[ R^{(\a)}(\a,\mu) + R^{(\a)}(\a,-\mu) \big] (h-\la^2) \cdot \f{ \dd \mu }{2\pi} \;. 
\enq
Clearly, the first two terms are strictly positive and behave at least as $\a^2$ when $\a \tend \infty$. 
It follows from the bounds obtained in lemma \ref{Lemme existence dvpm series det et resolvent} 
that the last term is, at most a $\e{O}(\a)$. Thence, $\veps_0(\a \mid \a) \limit{\a}{+\infty} +\infty$. 
The function $\a \mapsto \veps_0(\a,\a)$ being continuous, it admits at least one zero on $\intoo{0}{+\infty}$ by virtue of the
intermediate value theorem. 

We have already established that $(\la,\a) \mapsto \veps_0(\la\mid \a)$ is $\mc{C}^1$. This adjoined to 
\eqref{ecriture derivee eps0 alpha}-\eqref{ecriture derivee eps0 lambda} and Lemma \ref{Lemme existence dvpm series det et resolvent}
ensures that the partial derivatives $\Dp{\la}\veps_0$ and $\Dp{\a}\veps_0$ are in fact $\mc{C}^1$, \textit{ie} that, in fact, 
$\veps_0$ is $\mc{C}^2$. Hence, by induction, one gets that $\veps_0$ is smooth. 
In particular, the map $\a \mapsto \veps(\a\mid \a)$ is smooth. Since $[\dd \veps(\a \mid \a) / {\dd \a } ]_{ \a = q} >0$,
by the local inversion theorem for smooth functions, 
there exists open neighbourhoods $O_q$ of $q$ and $V_0$ of $0$ in $\R$ such that $\veps(\a \mid \a)$ is a
$\mc{C}^{\infty}$-diffeomorphism from $O_q$ onto $V_0$. \qed



\section{The low-temperature expansions}
\label{Section low-tempe}

In this section, building on the results of the previous two sections, we prove Theorem \ref{Theorem existence of AE for YangYang energy}. 
We start by proving the existence of the asymptotic expansion at order $0$, which represents the hardest part of the 
theorem. Then, by using a convenient rewriting of Yang-Yang's equation \eqref{ecriture eqn YY}, we show that 
the knowledge of the existence of an asymptotic expansion up to order $2\ell+2$ as given in \eqref{Theorem ecriture DA lowT veps}
guarantees the validity of this asymptotic expansion \eqref{Theorem ecriture DA lowT veps} to the order $2\ell+4$.

\subsection{Asymptotic expansion to order $0$}

Theorem  \ref{Theorem Yang-Yang existence solutions} guarantees that the unique 
solution to the Yang-Yang equation is entire and has a unique zero $\wh{q}$ which, in virtue of 
Proposition \ref{Proposition bounds zero veps}, satisfies $\wh{q} \geq \sqrt{h} $ uniformly in $T$. 
Since $\veps^{\prime}\!(\la) > 2\la$ for $\la >0$, it follows that 
 $| \veps^{\prime}\!( \, \wh{q}\, ) | > 2 \sqrt{h}$ uniformly in $T$. 
As a consequence, by the Bloch-Landau theorem (see \textit{eg} \cite{DemaillyAnalyseComplexe}), there exists an open neighbourhood 
\beq
U_{\wh{q}} \subset \mc{D}_{\wh{q}, \sqrt{h}/2} 
\qquad \e{and} \; \e{a} \; \e{radius} \qquad \de \geq \f{ 1 }{22} \cdot \f{ \sqrt{h} }{ 2 }  \cdot |\veps^{\prime}(\wh{q})| > \f{ h }{22}  
\qquad \e{such} \; \e{that} \qquad   \veps : U_{\wh{q}} \mapsto \mc{D}_{0,\de} 
\enq
is a biholomorphism. Above, $\mc{D}_{a,\eta}$ stands for
the open disk of radius $\eta$ centred at $a$. 
In the following, we shall write $\veps^{-1}$ for the associated local inverse of $\veps$, 
\textit{ie} $\veps^{-1}(0)=\wh{q}$. 
 This being established, we continue by decomposing
\beq
 - T \Int{\R}{}  K(\la-\mu) \ln\Big[ 1 + \ex{-\f{ \veps(\mu) }{ T } } \Big] \cdot \f{\dd \mu}{2\pi} \;  = \; 
  \Int{ -\wh{q} }{ \wh{q} } \veps(\mu) K(\la-\mu) \f{ \dd \mu  }{ 2\pi } \; + \;  \mc{V}^{(0)}(\la)  \; + \;   \mc{V}^{(\infty)}(\la) 
\nonumber
\enq
where
\beq
\mc{V}^{(\infty)}(\la) \; = \;  - T \Int{ J^{\de} }{} \hspace{-1mm} K(\la-\mu) 
						\ln\Big[  1 + \ex{-\f{ |\veps(\mu)| }{T} } \Big]  \cdot \f{\dd \mu}{2\pi}
\qquad \e{and} \qquad 
\mc{V}^{(0)}(\la)  \; = \; 
- T  \Int{ J^{\de}_{\wh{q}}\cup J^{\de}_{-\wh{q}} }{} \hspace{-2mm}  K(\la-\mu) \ln\Big[  1 + \ex{-\f{ |\veps(\mu)| }{T} } \Big] \cdot \f{ \dd \mu }{ 2\pi }\;. 
\enq
There, we have defined $J^{\de}_{\wh{q}}  = \veps^{-1} \big(\intoo{-\de}{\de} \big)$, 
$J^{\de}_{-\wh{q}}  = -\veps^{-1} \big(\intoo{\de}{-\de} \big)$ and we agree upon
 $J^{\de} = \R \setminus \big\{ J^{\de}_{\wh{q}}\cup J^{\de}_{-\wh{q}} \big\} $. Note that both $J_{\wh{q}}^{\de}$ 
and  $J_{-\wh{q}}^{\de}$  have a positive orientation. 

Since $\veps$ is even and strictly increasing on $\intoo{0}{+\infty}$, it follows that $|\veps(\la)| \geq \de$ for any 
$\la \in J^{\de}$. As a consequence, for any $\la \in \R$, one has 
\beq
\big| \mc{V}^{(\infty)}(\la) \big|  \; \leq \;   T \ln\Big[  1 + \ex{-\f{ \de  }{T}} \Big] \times 
\Int{ \R  }{} \!\!   K(\la-\mu)  \cdot  \f{ \dd \mu }{ 2\pi}  \; \leq \;  T \ln\Big[  1 + \ex{-\f{ \de  }{T} } \Big] \;. 
\enq
Thus $\mc{V}^{(\infty)}(\la)$ only generates exponentially small corrections in $T$. 
The fact that $\veps$ is even allows us to reduce the integration in $\mc{V}^{(0)}$ to $J_{\wh{q}}^{\de}$ only.
Then, the change of variables $\mu=\veps^{-1}(s)$ 
leads to 
\beq
\mc{V}^{(0)}(\la) \;  = \; 
-  T \Int{-\de}{\de} \bigg\{ \f{K(\la- \veps^{-1}(s) ) + K(\la + \veps^{-1}(s) ) }{\veps^{\prime}\circ \veps^{-1}(s)} \bigg\}  
\ln\Big[ 1+\ex{-\f{ |s| }{T}} \Big] \cdot \f{ \dd s }{ 2\pi }  \;.
\enq
Since 
\beq
\inf_{s \in \intoo{-\de}{\de} } \big| \veps^{\prime}\circ\veps^{-1}(s) \big|  \; \geq  \; 
\inf_{s \in \big] \,  \wh{q}-\f{\sqrt{h}}{2} \,  ; \,\wh{q} + \f{\sqrt{h}}{2} \,   \big[ } \hspace{-5mm} | \veps^{\prime}(s) | 
\; \geq  \; \inf_{s \in \big]\,  \f{\sqrt{h}}{2} \, ; \,  +\infty \, \big[ } \hspace{-5mm} | \veps^{\prime}(s) |  
\;  \geq  \; \sqrt{h}\; ,
\enq
it follows that, for any $\la \in \R$, 
\beq
\big| \mc{ V }^{(0)}(\la) \big|  \; \leq \;  4 T^2 \f{\norm{ K }_{ L^{\infty}(\R) } }{ 2\pi \sqrt{h}} 
\Int{ 0 }{ +\infty } \ln\Big[ 1+\ex{-s } \Big] \dd s  \; =  \; 
 \f{ T^2  \pi   }{ 3 c \sqrt{h} }   \;. 
\label{ecriture estimation grossiere R0}
\enq
Hence, the Yang-Yang equation can be recast in the form  
\beq
\veps(\la) \; = \;  \la^2 \; - \; h  \; + \;  \Int{ - \wh{q} }{  \wh{q} }
\hspace{-1mm}  K(\la-\mu) \veps(\mu) \cdot \f{ \dd \mu }{ 2\pi } 
 \; + \; \mc{V}^{(0)}(\la) \; + \; \mc{V}^{(\infty)}(\la) 
\label{ecriture representation NLIE veps sous forme prete a DA}
\enq
with the remainders  $ \mc{V}^{(0)}(\la) \; + \; \mc{V}^{(\infty)}(\la) \; = \;  \e{O}(T^2) $ 
being understood in the $ L^{\infty}(\R)$ sense. By Proposition \ref{Proposition bounds zero veps} 
there exist constants $w$ and $T_0$ such that $\wh{q}\leq 2 \sqrt{w}$  uniformly in $T \in \intff{0}{T_0}$. 
It thus follows from Lemma \ref{Lemme existence dvpm series det et resolvent} that $I-\tf{K}{(2\pi)}$
understood as an operator on $L^{2}\big(\intff{-\wh{q}}{\wh{q}}\big)$ is invertible with inverse
$I + \tf{R^{(\, \wh{q} \, )}}{ (2\pi) }$. Moreover, as $\wh{q}$ is  bounded  for $T \in \intff{0}{T_0}$, 
it follows from \eqref{ecriture bornes resolvent sur interval general} that 
the resolvent kernel $R^{(\, \wh{q} \, )}(\la,\mu)$ is bounded on $\R^2$ by a $T$ independent
constant. 

Thus recalling that $\veps_0(\la\mid \wh{q})$ refers to the unique solution to equation \eqref{equation int veps alpha} 
(\textit{cf} Proposition  \ref{Proposition preuve existence q pour epsilon 0}),  
it follows that 
\beq
\veps(\la)  \; = \; \veps_0(\la \mid \wh{q}) \; + \;   \mc{W}^{(0)}(\la) \; + \; \mc{W}^{(\infty)}(\la) \;. 
\enq
The last two function are given by 
\beq
\mc{W}^{(u)}(\la)   \; =\;  \mc{V}^{(u)}(\la) 
\; + \;  \Int{-\wh{q} }{ \wh{q} } R^{(\, \wh{q}\, )}(\la, \mu) \mc{V}^{(u)}(\mu) \cdot \f{\dd \mu }{2\pi } \qquad \e{with} 
\; \e{either} \qquad u =0 \; \e{or} \; u=\infty \;. 
\enq
As the resolvent kernel $R^{(\, \wh{q} \, )}(\la,\mu)$ is bounded on $\R^2$ by a $T$ independent
constant, it follows that  
\beq
\norm{ \mc{W}^{(0)} }_{ L^{\infty}(\R) } \; = \; \e{O}(T^2) \qquad \e{ whereas } \qquad    
\norm{ \mc{W}^{(\infty)} }_{ L^{\infty}(\R) } \; = \; \e{O}( \ex{-\f{ \de }{T} }) = \e{O}(T^{\infty})\;. 
\enq
 This means that the zero $\wh{q}$ of $\veps(\la)$ solves the equation 
\beq
\veps(\wh{q}) \;= \;  \veps_0( \wh{q} \mid \wh{q}) \; + \; \e{O}(T^2)  \;  = \;   0 \;. 
\enq
It has been shown in Proposition \ref{Proposition preuve existence q pour epsilon 0} that there exists a unique
$q \in \intoo{0}{+\infty}$ 
such that $\veps_0(q\mid q)=0$. Moreover, there also exist open neighborhoods $O_q$ of $q$ and $V_0$ of $0$ in $\R$
such that $m: \a \mapsto \veps_0(\a\mid \a)$ is a $\mc{C}^{\infty}$-diffeomorphism from $O_q$ onto $V_0$. 
Since for T small enough, one has $\e{O}(T^2) \in V_0$, it follows that  $\wh{q}=  m^{-1}( -\e{O}(T^2) )$
is well defined. Moreover, the Taylor expansion of $m^{-1}$ around $0$ implies that 
$\wh{q} = q \; + \; \e{O}(T^2)$.

Note that one has the decompositon
\beq
\Int{ -\wh{q} }{ \wh{q} } \veps(\mu) K(\la-\mu) \cdot \f{ \dd \mu  }{ 2\pi }   \; =  \; 
\Int{ -q }{ q }  \!  \veps(\mu) K(\la-\mu) \cdot  \f{ \dd \mu  }{ 2\pi }  \; + \;
 \Int{q }{ \wh{q} } \! \veps(\mu) \big[ K(\la-\mu)+ K(\la + \mu) \big]\cdot  \f{ \dd \mu  }{ 2\pi } \;. 
\label{ecriture approximation integrale avec q hat vers integrale avec q}
\enq
We have already established that $\veps(\la) = \veps_0(\la \mid \wh{q} ) \; + \; \e{O}(T^{2})$ with the $\e{O}(T^2)$
being uniform in $\la \in \R$. Since $\veps_0(\la \mid \wh{q})$ is uniformly bounded in some neighbourhood 
containing $q$ and $\wh{q}$, knowing that $\wh{q} = q \; +  \;   \e{O}(T^2)$ ensures that 
the second term in \eqref{ecriture approximation integrale avec q hat vers integrale avec q}
is a $\e{O}(T^2)$, again in the $L^{\infty}(\R)$ sense. This means that, in fact, $\veps(\la)$ satisfies 
\beq
\veps(\la) \; = \;  \la^2 \; - \; h  \; + \;  \Int{ - q }{  q } K(\la-\mu) \veps(\mu) \f{ \dd \mu }{ 2\pi } 
 \; + \; \wt{\mc{V}}^{(0)}(\la) \; + \; \mc{V}^{(\infty)}(\la)   \; ,
\label{equation integrale pour eps forme parfaite pour asymptoptique}
\enq
where 
\beq
\wt{\mc{V}}^{(0)}(\la) \; = \; \mc{V}^{(0)}(\la) \; + \;
 \Int{q }{ \wh{q} } \veps(\mu) \big[ K(\la-\mu)+ K(\la + \mu) \big]\cdot  \f{ \dd \mu  }{ 2\pi } \;. 
\label{definition V tilde}
\enq
Again,  $\wt{\mc{V}}^{(0)}(\la) \; + \; \mc{V}^{(\infty)}(\la) \; = \; \e{O}(T^2)$ in the $L^{\infty}(\R)$ sense. Thus, 
by acting with the operator $I+\tf{R^{(q)}}{2\pi}$ on both sides of
\eqref{equation integrale pour eps forme parfaite pour asymptoptique} 
and using, again, the estimates 
\eqref{ecriture bornes resolvent sur interval general}  we get that 
$\veps(\la) = \veps_0(\la \mid q ) \; + \; \e{O}(T^{2})$. 
The analytic properties of the resolvent kernel $R^{(q)}(\la,\mu)$ obtained in Proposition 
\ref{Proposition preuve existence q pour epsilon 0} guarantee that, given any $a \in \intoo{0}{\tf{c}{2}}$,
$\veps_0(\la \mid q)$ is indeed analytic on the strip $\mc{S}_{a}$.   
As $\veps$ is entire, it thus follows that the remainder must also be holomorphic on $\mc{S}_a$. 
Finally, it is readily seen from its explicit expression that the remainder is stable under finite
order $\la$-differentiations. This last result can also be derived from Cauchy's integral representation using 
the fact that the remainder is holomorphic and uniform.

\subsection{Existence to all orders}

We now establish the existence of the asymptotic expansion to all orders in $\ell$ by induction. 
Thus assume that equation \eqref{Theorem ecriture DA lowT veps} given in Theorem \ref{Theorem existence of AE for YangYang energy}
holds for some $\ell$.

Observe that, in some vicinity of $ 0 $, the local inverse  $\veps^{-1}(\la)$ of $\veps(\la)$ 
admits the integral representation  
\beq
 \veps^{-1}(\la) = \Oint{  \Dp{}U_{\wh{q}} }{}   \f{ z \; \veps^{\prime}(z)  }{ \veps(z) - \la }  \f{ \dd z  }{2i\pi}
\qquad \e{for} \qquad \la \in \mc{D}_{0,\tf{\de}{2}} \;. 
\label{ecriture rep int inverse veps}
\enq
Note that, for any $z \in \Dp{}U_{\wh{q}}$ 
one has $\veps(z) \in \Dp{}\mc{D}_{0,\de}$. Thus, for any $\la \in \mc{D}_{0,\tf{\de}{2}}$
one has $ | \veps(z) - \la | \geq \tf{\de}{2}$ so that the integral representation \eqref{ecriture rep int inverse veps}
is indeed well-defined. Moreover, this representation is well fit for providing a low-$T$ asymptotic expansion 
for the function $\veps^{-1}$ in the neighbourhood of $0$ as soon as the one for $\veps$ is known. 
In particular, it implies that $\veps^{-1}$ admits a low-$T$ asymptotic expansion to the same order as $\veps$.

So as to push the asymptotic expansion of $\veps(\la)$ two orders further, 
by acting with $I+R^{(q)}/(2\pi)$ on both sides of \eqref{equation integrale pour eps forme parfaite pour asymptoptique},
we obtain the representation
\beq
\veps(\la) \; = \; \veps_0(\la \mid q)  \; + \; \mc{I}^{(1)}(\la) \; + \; \mc{I}^{(2)}(\la) \; + \; 
\mc{I}^{(\infty)}(\la) \;, 
\enq
where 
\beq
 \mc{I}^{(1)}(\la) \; = \; -   \;  T \Int{-\de}{\de} \bigg\{ \f{\ov{R}(\la, \veps^{-1}(s)  ) }{\veps^{\prime}\circ \veps^{-1}(s)} \bigg\}  
\ln\Big[ 1+\ex{-\f{ |s| }{T}} \Big] \cdot  \f{ \dd s }{ 2\pi }   \qquad \e{where} \qquad 
\ov{R}(\la,\mu) \; = \; R^{(q)}(\la,\mu) \, + \, R^{(q)}(\la, - \mu)  \;, 
\enq
and
\beq
\mc{I}^{(2)}(\la) \; = \;  \Int{q }{ \wh{q} } \veps(\mu) \cdot \ov{R}(\la,\mu)  \cdot \f{ \dd \mu  }{ 2\pi }  \; 
\qquad  \qquad 
\mc{I}^{(\infty)}(\la) \; = \;  - T \Int{ J^{\de} }{}  R^{(q)}(\la,\mu) \ln\Big[  1 + \ex{-\f{ |\veps(\mu)| }{T} } \Big] 
\cdot \f{\dd \mu}{2\pi} \;. 
\enq
Moreover, below, we fix a particular value of $a$, $0<a <\tf{c}{2}$ defining the width of the strip $\mc{S}_{a}$. 
First, we obtain the asymptotic expansion of $\mc{I}^{(1)}(\la)$. 
It follows from the Taylor integral expansion 
\beq
\f{\ov{R}(\la, \veps^{-1}(s)  ) }{\veps^{\prime}\circ \veps^{-1}(s)}
\; = \;  \sul{r=0}{2\ell + 1}  \f{s^r}{r!} \f{ \Dp{}^r }{ \Dp{}s^{r} } 
\bigg\{ \f{\ov{R}(\la, \veps^{-1}(s)  ) }{\veps^{\prime}\circ \veps^{-1}(s)} \bigg\}_{\mid s=0}
\; + \; g_{2\ell + 2}(\la, s ) 
\nonumber
\enq
with 
\beq
 g_r(\la,s) \; = \; \f{ s^{r} }{ (r-1)! } \Int{0}{1}
 (1-u)^{r-1} \cdot \f{ \Dp{}^r }{ \Dp{}s^{r} } 
 \bigg\{ \f{\ov{R}(\la, \veps^{-1}(s)  ) }{\veps^{\prime}\circ \veps^{-1}(s)} \bigg\}_{\mid s=tu}  \hspace{-2mm}\dd u 
\nonumber
\enq
that one has the representation 
\beq
\mc{I}^{(1)}(\la) \; = \;  - \sul{r=0}{ \ell }  \f{T^{2(r+1)}}{(2r)!} \f{ \Dp{}^{2r} }{ \Dp{}s^{ 2r } } 
\bigg\{ \f{\ov{R}(\la, \veps^{-1}(s)  ) }{\veps^{\prime}\circ \veps^{-1}(s)} \bigg\}_{ \mid s=0}
\Int{-\f{\de}{T} }{ \f{\de}{T} } s^{2r} \ln \big[1+\ex{-|s|} \big] \cdot \f{ \dd s }{2\pi}
\; - \; T \Int{-\de }{ \de } g_{ 2 \ell + 2 }(\la, s ) \ln \big[1+\ex{ - \f{|s|}{T} } \big] \cdot \f{ \dd s }{ 2\pi }   \;. 
\nonumber
\enq

As the asymptotic expansion of $\veps$ is uniform with respect to differentiations and $\ov{R}$
is smooth with bounded derivatives on the strip $\ov{\mc{S}}_a \times \ov{\mc{S}}_a$, it follows that 
\beq
|g_{2\ell + 2}(\la,s)| \; \leq  \; s^{2 \ell + 2} C_{\ell}  \qquad  
\e{uniformly} \; \e{in} \quad \la \in \ov{\mc{S}}_{a} \quad \e{and} \quad s \in \intff{-\de}{\de}
\enq
where $C_{\ell}$ is some $T$-independent constant.  Hence
\beq
\Big| T  \Int{-\de }{ \de } g_{2\ell + 2}(\la, s ) \ln \big[1+\ex{-\f{|s|}{T}} \big] \cdot \dd s   \Big| \; \leq  \; 
2 C_{\ell} T^{2\ell + 4} \Int{0}{+\infty} u^{2\ell + 2}\ln [ 1+\ex{-u} ] \cdot \dd u \; = \; \e{O}(T^{2 \ell + 4 }) 
\enq
Furthermore, up to exponentially small corrections in $T$, one can replace the integration over $\intff{-\tf{\de}{T}}{\tf{\de}{T}}$
with one over $\R$. The corrections will, again, be analytic, uniform in $\la \in  \ov{\mc{S}}_{a}$
and stable under finite order differentiation. 
Hence, using the integral representation for the Riemann $\zeta$-function
\beq
\big( 1-2^{-1-r}\big) \zeta\pa{r+2} = \f{1}{\Ga\pa{r+1}} \Int{0}{+\infty} t^{r} \ln\pa{1+\ex{-t}} \dd t \; ,
\enq
we get 
\beq
\mc{I}^{(1)}(\la) \; = \;  - \sul{r=0}{\ell }  2 T^{2(r+1)} \big( 1-2^{-1-2r}\big) \cdot \zeta(2+2r) \cdot \f{ \Dp{}^{2r} }{ \Dp{}s^{2r} } 
\bigg\{ \f{\ov{R}(\la, \veps^{-1}(s)  ) }{\veps^{\prime}\circ \veps^{-1}(s)} \bigg\}_{s=0}
\; + \; \e{O}(T^{2(\ell + 2)}) \;. 
\enq
By composition of the Taylor expansion of $\ov{R}(\la,\mu)$ in $\mu$ belonging to a neighborhood of $\wh{q}$, 
and of the asymptotic expansions of $\veps^{-1}(s)$ and $\tf{1}{ \big( \veps^{\prime}\circ \veps^{-1}(s) \big) } $ 
into powers of $T$ one readily gets that there exist functions $r_p^{(1)}(\la)$, $p=1,\dots, \ell$ analytic in the strip 
$\mc{S}_{ a }$ such that 
\beq
\mc{I}^{(1)}(\la) \; = \;  - \sul{r=0}{ \ell }   T^{2(r+1)} r_p^{(1)}(\la) 
\; + \; \e{O}(T^{2(\ell + 2)}) \;. 
\enq
Again, the remainder is stable in respect to finite order differentiations, uniform and analytic in $\mc{S}_a$.

Clearly, $\mc{I}^{(\infty)}(\la) \; = \; \e{O}(T^{\infty})$, with a remainder that enjoys the usual properties. 
Thus, it remains to estimate $\mc{I}^{(2)}(\la)$. 
Inserting the asymptotic expansion of $\veps(\la)$ 
up to $\e{O}(T^{2(\ell+1)})$ corrections leads to 
\beq
\mc{I}^{(2)}(\la) \; = \; 
\Int{q}{ \wh{q} }  \ov{R}(\la,\mu)  \cdot \veps_0(\mu\mid q) \f{ \dd \mu  }{2\pi}  
\; + \; \sul{p=1}{\ell} T^{2p} 
\Int{q}{ \wh{q} } \ov{R}(\la,\mu)  \cdot \veps_{2p}(\mu) \f{ \dd \mu  }{2\pi} 
\; + \; \e{O}(T^{2\ell+4}) \;. 
\label{ecriture developpement integrale bord a tout ordre}
\enq
We were able to replace the uniform $\e{O}(T^{2\ell +2})$ appearing in the asymptotic expansion of $\veps$
by a $\e{O}(T^{2\ell+4})$ since it has been integrated over an interval of length $|\wh{q}-q| = \e{O}(T^2)$. 
This $\e{O}(T^{2\ell+4})$ is differentiation stable, uniform and holomorphic in $\mc{S}_a$. 
Since $\wh{q}=\veps^{-1}(0)$, $\wh{q}$ admits a low-T asymptotic expansion 
to the same order as $\veps^{-1}$. The regularity of the integrands then ensures that 
there exist analytic functions $r_1^{(2)}(\la), \dots , r^{(2)}_{\ell+1}(\la)$ on $\mc{S}_{a}$ such that 
\beq
\sul{p=1}{ \ell } T^{2p} \Int{q}{ \wh{q} } \ov{R}(\la,\mu)  \cdot \veps_{2p}(\mu) \f{ \dd \mu  }{2\pi} 
\; = \; 
\sul{p=1}{ \ell } T^{2p} r_p^{(2)}(\la) \; + \;  \e{O}(T^{2 \ell +4}) \;. 
\label{ecriture DA rest integrale bord sans eps 0}
\enq
The remainder in \eqref{ecriture DA rest integrale bord sans eps 0}  is also readily checked to have the desired properties. 
Moreover, we do stress that one has gained a supplementary $T^2$ precision on the remainder since on the one hand the sum starts with a 
$T^2$ and on the other hand, one integrates over an interval of length $|\wh{q} - q | = \e{O}(T^2)$. 

The first integral in \eqref{ecriture developpement integrale bord a tout ordre} needs however a separate attention. 
Tayor expanding the integral to the order $(\wh{q}-q)^{ \ell +2}$ leads to 
\beq
\Int{q}{\wh{q}}  \ov{R}(\la,\mu) \cdot \veps_{0}(\mu\mid q) \cdot \f{\dd \mu}{2\pi}  \; = \; 
\sul{r=0}{\ell +1}\f{  (\wh{q}- q)^{r+1} }{(r+1)!}  \cdot \f{\Dp{}^{r}}{\Dp{}\mu^{r} } \bigg\{ 
\ov{R}(\la,\mu) \veps_{0}(\mu\mid q) 
 \bigg\}_{\mu= q } 
+ \; \e{O}\Big( (\wh{q}-q)^{ \ell +2} \Big) \;. 
\enq
Note that the $r=0$ terms of this expansion is zero since $\veps_0(q\mid q) \, = \, 0$. Thus, this expansion  
only contains terms $ (\wh{q}- q)^{r+1}$ with $r\geq 1$. Furthermore taking into account that $\wh{q}-q=\e{O}(T^2)$  and 
that $\wh{q}$ admits an asymptotic expansion to order $\e{O}(T^{2 \ell +2})$, it follows that 
$ (\wh{q}- q)^{r+1}$ with $r\geq 1$ admits an asymptotic expansion up to $\e{O}(T^{2 \ell +4})$, \textit{ie} 
with a $T^2$ better precision than $\wh{q}$. Again, the regularity properties of the integrand then guarantee that 
there exist holomorphic functions 
$r^{(3)}_2(\la), \dots, r_{\ell+1}^{(3)}(\la)$ on $\mc{S}_{a}$ such that 
\beq
\Int{q}{\wh{q}}  \ov{R}(\la,\mu) \cdot \veps_{0}(\mu\mid q) \cdot \f{\dd \mu}{2\pi}  \; = \; 
\sul{p=2}{ \ell +1} T^{2p} r_p^{(3)}(\la)   \; + \; \e{O}\Big( T^{2( \ell +2)} \Big) \;. 
\enq

Adding up together all of the asymptotic expansions that we have obtained leads to 
\beq
\veps(\la) \; = \; \veps_0(\la \mid q)  \; + \; 
\sul{p=1}{\ell +1} \Big( r_p^{(1)}(\la) \, + \,  r_p^{(2)}( \la ) \, + \,  r_p^{(3)}( \la ) \Big) \cdot T^{ 2p }
\; + \; \e{O}\Big( T^{2(\ell + 2)} \Big) \;. 
\enq
where we agree upon $r_1^{(3)}(\la)=0$. 
Hence, as claimed, $\veps(\la)$ admits a low-$T$ asymptotic expansion up to the order $\e{O}\Big( T^{2( \ell +2)} \Big)$
with coefficients being analytic functions of $\la$ on $\mc{S}_{a}$. 
Moreover, clearly, the remainder enjoys the claimed properties. \qed

\section*{Conclusion}

We have proposed a method allowing one to prove the existence of the low-temperature
expansion to all orders in $T$ of the unique solution of the Yang-Yang equation. 
In the course of the proof, we have also established the unique solvability of the 
system \eqref{ecriture equation veps GS et definition zero q} for the 
 dressed energy-endpoint of the Fermi zone couple $(\veps_0(\la\mid q), q)$ 
characterizing the zero temperature behaviour of the non-linear Schr\"{o}dinger model.  
Our method mainly utilizes the positivity of Lieb's kernel
so that it should provide one with analogous results for similar non-linear integral equations describing thermodynamics of other 
integrable models. Also, we expect that upon a few modifications, our techniques would allow one to 
approach rigorously the question of the large-volume behaviour of 
various thermodynamic quantities arising in integrable models, such as, for instance, 
the $\tf{1}{L}$ corrections to the energies of the ground and low-lying excited states.

\section*{Acknowledgements}

K.K.K. is supported by Centre National de la Recherche Scientifique. 
K.K.K. is grateful to the referees for their valuable comments.

\end{document}